\documentclass[a4paper,fleqn,usenatbib]{mnras}
\usepackage{amssymb,amsmath,graphicx,psfrag,mathtools}
\usepackage[T1]{fontenc} %MNRAS
\usepackage{ae,aecompl} %MNRAS
\usepackage{url}

\title[The effect of the variation of pulsar velocity fields]{On the effect of the variation of velocity fields in pulsars}
\author[DS]{Shuang Du$^{1}$\thanks{E-mail: dushuang@pku.edu.cn}\\
$^{1}${School of Mathematics and Computer, Tongling University, Tongling, Anhui, 244061, China}}

\date{\today}
\begin{document} %MNRAS
\label{firstpage} %MNRAS
\pagerange{\pageref{firstpage}--\pageref{lastpage}} %MNRAS
\maketitle %MNRAS
\begin{abstract}
Glitches are sudden spin-up events of pulsars and are usually thought to be induced by unpinning of neutron superfluid vortices in pulsar crusts.
Unpinning and repinning of superfluid vortices, and even thermoelectric effects induced by the deposited heat released during glitches, may vary the velocity fields in pulsars.
We show that the generally invoked magnetic dipole fields of pulsars cannot remain stationary during the variation of the velocity fields, so that multipole components must be generated.
We argue that the increase of the spark frequency of periodic radio pulses is the indicator for the emergence of the multipole components.
Interpretations of pulsar nulling, rebrightening of radio-quiet magnetars, differences between Crab and Vela pulsars after glitches, and extra-galactic fast radio burst-like events from SGR 1935+2154 have been proposed based on the influence of the variation of the velocity field on the magnetic field.
\end{abstract}
 %MNRAS
%\maketitle
\section{Introduction}
%MNRAS USES \citet IN PLACE OF \cite
The popular model for the generation of pulsar magnetic fields \citep{1993ApJ...408..194T} is based on the dynamo theory (see \citealt{1978mfge.book.....M} for more details).
In this scenario, convection and differential rotation convert the thermal energy and kinetic energy of fluids in pulsars into magnetic energy.
Therefore, if the dynamo theory is valid for the generation of pulsar magnetic fields, variations in the velocity field of charge particles in pulsars, in particular sudden variations, will inevitably induce the variation of pulsar magnetic fields. In general, the variation of the velocity field directly induced by the sluggish spindown due to equivalent magnetic dipole radiation should be too weak to give rise to observable phenomena during a short duration.
However, occasionally sudden spin-up events of pulsars (i.e., glitches; \citealt{1969Natur.222..228R,1969Natur.222..229R}; see \citealt{2015IJMPD..2430008H,2022Univ....8..641Z} for reviews)
provide the condition to investigate the effect of the variation of these interior velocity fields and to test the dynamo model of pulsar magnetic fields.
It has been wildly considered that unpinning of the neutron superfluid vortices in the pulsar crusts triggers these glitches
since repinning of these vortices is the only plausible way to explain post-glitch recovery processes (\citealt{1975Natur.256...25A}; see \citealt{2022NewA...9001655P} for another scenario).
Therefore, unpinning and repinning of superfluid vortices, and even thermoelectric effect induced by deposited heat due to glitches \citep{1996ApJ...457..844L},
should change the velocity fields in pulsars.
Glitches are observable phenomena, the after-effects associated with the resulted variation of these velocity fields may also be observable, especially for that of magnetars,
since, in comparison with normal radio pulsars, magnetars have more free energy to be released (e.g., \citealt{1995MNRAS.275..255T,1996ApJ...473..322T,2009ApJ...703.1044B})
as shown in the X-ray and gamma-ray bands (\citealt{1980ApJ...237L...7E,1998Natur.393..235K,2005Natur.434.1098H}; see \citealt{2017ARA&A..55..261K} for a review).

In Section \ref{sec2}, we discuss how the variation of the velocity field affects the magnetic field in a pulsar.
In Section \ref{sec3}, we show the indicator of the variation of the pulsar magnetic field.
In Section \ref{sec4}, we present several applications of the idea that glitches induce the variation of pulsar magnetic fields.
Section \ref{sec5} is the summary.

\section{Variation of magnetic fields}\label{sec2}
We aim to illustrate analytically the effect of the variation of the velocity field
and present the derivation under Newtonian gravity. According to Maxwell's equations, there is
\begin{eqnarray}\label{W8}
\mathbf{j}=\frac{1}{4\pi}\left ( \nabla\times\mathbf{B}-\frac{\partial \mathbf{E}}{\partial t} \right ),
\end{eqnarray}
where $\mathbf{j}$ is the current field, $\mathbf{B}$ is the magnetic field, $\mathbf{E}$ is the electric field and $t$ is the time.
The conductivities of pulsars, $\sigma$, are very large. One has $\mathbf{E}=-\mathbf{v}\times\mathbf{B}$ when $\sigma\rightarrow \infty$  \citep{1975clel.book.....J}.
Then, equation (\ref{W8}) can be rewrote as
\begin{eqnarray}\label{W9}
\mathbf{j}=\frac{1}{4\pi}\left ( \nabla\times\mathbf{B}+\frac{\partial \mathbf{(\mathbf{v}\times \mathbf{B}})}{\partial t} \right ),
\end{eqnarray}
where $\mathbf{v}$ is the velocity field of charge particles.
Note that, before the glitch, the pulsar is in a steady state (i.e., the initial velocity field, $\mathbf{v}_{0}$, magnetic field,
$\mathbf{B}_{0}$, and current field, $\mathbf{v}_{0}$, should be approximative to constants) and after the glitch the pulsar will ultimately recover to another steady state
$(\mathbf{v}_{0}+\delta\mathbf{v}, \mathbf{B}_{0}+\delta\mathbf{B}, \mathbf{j}_{0}+\delta\mathbf{j})$.
In both of the two steady states, there should be
\begin{eqnarray}
\frac{\partial \mathbf{E}}{\partial t}=-\frac{\partial \mathbf{(\mathbf{v}\times \mathbf{B}})}{\partial t}=0,
\end{eqnarray}
due to the large conductivity ($\partial \mathbf{B}/\partial t =0$; see equation \ref{W1})
and sluggish spindown ($\partial \mathbf{v}/\partial t =0$).
This means that displacement current vanishes under the steady state and only conduction current is left.
Therefore, according to equation (\ref{W9}), we have
\begin{eqnarray}\label{W10}
\mathbf{j}=n_{\rm e}\mathbf{v}=\frac{1}{4\pi}\left ( \nabla\times\mathbf{B}\right ),
\end{eqnarray}
where $n_{\rm e}$ is the net charge density.

On the other hand, since the conductivity is very large, and the evolution of the pulsar magnetic field is slow and secular, the induction equation in a short duration is given by
\begin{eqnarray}\label{W1}
\frac{\partial \mathbf{B}}{\partial t}=\nabla\times (\mathbf{v}\times \mathbf{B})=0.
\end{eqnarray}
Combining equations (\ref{W10}) and (\ref{W1}), we get
\begin{eqnarray}\label{W11}
\nabla\times[(\nabla\times \mathbf{B})\times\mathbf{B}]=0.
\end{eqnarray}
Under the spherical coordinate frame $(r, \theta, \varphi)$ and the assumption of axisymmetric magnetic field,
equation (\ref{W11}) can be decomposed into
\begin{eqnarray}\label{W12}
&&(\nabla_{\theta}B_{\rm r})(\nabla_{\rm r}B_{\varphi})+B_{\rm r}\nabla_{\theta}\nabla_{\rm r}B_{\varphi}\nonumber\\
&+&(\nabla_{\theta}B_{\theta})(\nabla_{\theta}B_{\varphi })+B_{\theta}\nabla_{\theta}\nabla_{\theta}B_{\varphi}=0,
\end{eqnarray}
\begin{eqnarray}\label{W13}
&&(\nabla_{\rm r}B_{\rm r})(\nabla_{\rm r}B_{\varphi})+B_{\rm r}\nabla_{\rm r}\nabla_{\rm r}B_{\varphi}\nonumber\\
&-&(\nabla_{\rm r}B_{\theta})(\nabla_{\theta}B_{\varphi })-B_{\theta}\nabla_{\rm r}\nabla_{\theta}B_{\varphi}=0,
\end{eqnarray}
and
\begin{eqnarray}\label{W14}
&&(\nabla_{\theta}B_{\theta})(\nabla_{\rm r}B_{\theta})+B_{\theta}\nabla_{\theta}\nabla_{\rm r}B_{\rm \theta}\nonumber\\
&-&(\nabla_{\theta}B_{\theta})(\nabla_{\theta}B_{\rm r})-B_{\theta}\nabla_{\theta}\nabla_{\theta}B_{\rm r}\nonumber\\
&-&(\nabla_{\rm r}B_{\rm r})(\nabla_{\theta }B_{\rm r})-B_{\rm r}\nabla_{\rm r}\nabla_{\theta}B_{\rm r}\nonumber\\
&+&(\nabla_{\rm r}B_{\rm r})(\nabla_{\rm r}B_{\theta })+B_{\rm r}\nabla_{\rm r}\nabla_{\rm r}B_{\theta }=0,
\end{eqnarray}
where $\nabla_{\rm r}=\frac{\partial}{\partial r}$ and $\nabla_{\theta}=\frac{1}{r}\frac{\partial}{\partial \theta}$.
It is easy to see that the usually invoked magnetic dipole field,
\begin{eqnarray}\label{ap1}
\left\{\begin{matrix}
B_{\rm r}\propto r^{-3}\cos \theta \\
B_{\theta}\propto \frac{1}{2}r^{-3}\sin \theta
\end{matrix}\right.,
\end{eqnarray}
cannot satisfy equation (\ref{W14}).
According to the symmetry of the three subscripts and the passivity of magnetic field that $\nabla\cdot\mathbf{B}=0$, one can find some particular solutions.
For example, the following solution, which has the similar mirror symmetry with respect to the equatorial plane to that of the magnetic dipole field,
\begin{eqnarray}\label{eq11}
\left\{\begin{matrix}
B_{\rm r}= \left (\eta C_{1} -{\xi}\ln r\right )\cos \theta \\
B_{\theta}= \left (\xi C_{1}- \frac{\xi^{2} }{\eta}\ln r \right )\sin \theta\\
B_{\varphi}=C
\end{matrix}\right.,
\end{eqnarray}
can satisfy equation (\ref{W11}), where $\xi$, $\eta$, $C_{1}$ and $C$ are constants.

Note that, the above discussion do not invoke the variation induced by the glitch,
so equation (\ref{eq11}) is at least valid beyond superconduction region of protons of the pulsar, as long as the pulsar magnetosphere is also static (but, another set of values of $\xi$, $\eta$, $C_{1}$ and $C$ is needed).

If we consider the change resulted by the glitch, more constraints will arise.
Let us consider the perturbation that $\mathbf{v_{0}}\rightarrow\mathbf{v_{0}}+\delta\mathbf{v}$, $\mathbf{B_{0}}\rightarrow \mathbf{B_{0}}+\delta\mathbf{B}$,
$\mathbf{j}_{0}\rightarrow \mathbf{j}_{0}+\delta \mathbf{j}$. Inevitably, the pressure $\mathcal{P}$,
mass density $\rho$, and gravitational potential $\phi$, will be also changed (i.e., the right hand side of equation \ref{W4}).
According to equation (\ref{W1}) and Euler Equation that
\begin{eqnarray}\label{W2}
\frac{\mathbf{j}}{c}\times \mathbf{B}=\nabla \mathcal{P}+\rho \nabla\phi,
\end{eqnarray}
we get
\begin{eqnarray}\label{W3}
\nabla\times[\mathbf{v}_{\rm 0}\times\delta\mathbf{B}+\delta\mathbf{v} \times(\mathbf{B}_{\rm 0}+\delta\mathbf{B})]=0,
\end{eqnarray}
and
\begin{eqnarray}\label{W4}
[\mathbf{j}_{\rm 0}\times\delta \mathbf{B}+\delta \mathbf{j}\times(\mathbf{B}_{\rm 0}+\delta \mathbf{B})]=\delta[\nabla \mathcal{P}+\rho \nabla\phi].
\end{eqnarray}
Substituting equations (\ref{W10}) and (\ref{W3}) into equation (\ref{W4}), we get
\begin{eqnarray}\label{W5}
&&(\nabla\times n_{\rm e})[\mathbf{v}\times \delta \mathbf{B}+\delta\mathbf{v}\times(\mathbf{B}+\delta \mathbf{B}) ]\nonumber\\
&+&\nabla\times [\delta n_{\rm e}\mathbf{v}\times(\mathbf{B}+\delta \mathbf{B})]
=\nabla\times\delta[\nabla \mathcal{P}+\rho \nabla\phi].
\end{eqnarray}
Considering that the perturbation with the same order vanishes, equation (\ref{W5}) reduces to
\begin{eqnarray}\label{W6}
\nabla\times \delta (n_{\rm e}\mathbf{v}\times\mathbf{B})=\nabla\times\delta[\nabla \mathcal{P}+\rho \nabla\phi],
\end{eqnarray}
and
\begin{eqnarray}\label{W7}
(\nabla n_{\rm e})(\delta \mathbf{v}\times \delta\mathbf{B})+\nabla\times(\delta n_{\rm e}\mathbf{v}\times \delta\mathbf{B})=0.
\end{eqnarray}
According to the gravitational-wave observations of Galactic pulsars (note that pulsars are oblique rotors), the ellipticity of these pulsars is very small \citep{2014ApJ...785..119A,2017PhRvD..96f2002A}.
This indicates that the distribution of the magnetic field has little effect on the overall mass distribution in a pulsar.
Therefore, the right hand side of equation (\ref{W6}) should be actually higher-order perturbation when compares with the left hand side of equation (\ref{W6}).
Then, $\delta (n_{\rm e}\mathbf{v}\times\mathbf{B})$ should be the gradient of a scalar field.
This is an unnatural requirement since the magnetic field is axisymmetric (see equations (\ref{ap1}) and (\ref{eq11})), and the other two parameters need to be fine tuned.
Actually, by comparing equation (\ref{W3}) with equation (\ref{W4}) regardless of equations (\ref{W6}) and (\ref{W7}),
one can see that a special condition can maintain equation (\ref{W4}) under a generic magnetic field: all the perturbations of scalar fields are uniform, as well as the number density of charge particles.
Of course, this is too idealized.

According to the above discussion, we find that,
(a) the usually invoked magnetic dipole field can not keep steady even in the presence of the toroidal field;
(b) when a perturbation of the fluid arises in a pulsar, only the magnetic field with very special structure can keep steady.
Thus, we conclude that the multipole component of the magnetic field must be generated when a glitch occurs.
It is worth noting that since multiple components possess smaller scales with respect to the dipole component, the apparent variation of the magnetic field on the pulsar should be localized.

\section{The indicator of the variation of the magnetic field}\label{sec3}

Although the radiation mechanism of pulsed radio emission of pulsars is not well understood,
it is widely believed that the radiation mechanism is closely related to the structure of the magnetosphere \citep{1975ApJ...196...51R,1979ApJ...231..854A,1987ApJ...320..333U,2008ApJ...683L..41B,2020PhRvL.124x5101P}.
Therefore, if the newborn multipole component arise from the open field line region,  the properties of the pulsed radio emission should be changed.
In the following, we employ the popular inner gap model \citep{1975ApJ...196...51R} to illustrate this issue,
since models invoking gap-like accelerators for pulsar radio emission (include the model shown in \citealt{1979ApJ...231..854A}) are able to correctly account for the death line shown in the two-dimensional pulsar parameter phase space (e.g., $P-\dot{P}$ diagram,
where $P$ is the pulsar period and $\dot{P}$ is the pulsar spin-down rate \citealt{1993ApJ...402..264C,2000ApJ...531L.135Z}).

Let us first briefly review the inner gap model.
For a pulsar with net positive charges being in its open field line region,
when these positive charges move outwards \citep{1971ApJ...164..529S,1973NPhS..246....6H},
the lost charges can not be replenished by the stellar surface due to large binding energy of positive charges.
Whereafter, a gap grows on the polar cap.
The potential across the gap, $U$, increases with the gap height, $h$, initially.
When the voltage increases to a critical value (inversely proportional to the curvature radius of magnetic field lines),
positrons (e.g., originally produced by the thermal photons from the pulsar surface through $\gamma+\mathbf{B}\rightarrow e^{-}+e^{+}+\mathbf{B}$)
in the gap can be accelerated to a high energy that the photons emitted by these positrons via curvature radiation and even inverse Compton scattering \citep{1998A&A...333..172Q}
will again be converted into electron-positron pairs via the reaction $\gamma+\mathbf{B}\rightarrow e^{-}+e^{+}+\mathbf{B}$.
Avalanches of discharges lead to the eventual reduction of the potential, after which the spark ceases.
The growth and reduction of the gap happens back and forth, generating a large number of secondary charges to sustain the pulsed radio emission.
However, the maximum value of the potential across the gap does not grow endlessly with the gap height,
but is enslaved by the pulsar period (the potential is inversely proportional to the period).
There is a moment when the maximum value of the potential cannot sustain a spark due to the spindown of the pulsar, and then the pulsed radio emission extinguishes.

The curvature radius of magnetic field lines of the multipole component is smaller than that of the dipole component.
Therefore, under the inner gap model, the spark frequency should increase when magnetic multipole component emerges from the polar cap.
Here, an equivalent model is presented to illustrate this issue.

We note that the electric field across the gap behaves like that of a parallel-plate capacitor since the gap height is much smaller than the radius of the polar cap, $r_{\rm p}$.
For example, the electric field of the parallel-plate capacitor is
\begin{eqnarray}
{E}'=\frac{{U}'}{{d'}}=\frac{4\pi k Q'}{S'}
\end{eqnarray}
where $U'$ and $d'$ are the voltage and interval between the two plates, $Q'$ and $S'$ are the charge and area of each plate, and $k$ is the electrostatic constant.
As a contrast, the electric field across the gap is \citep{1975ApJ...196...51R}
\begin{eqnarray}\label{e2}
E\approx \frac{2U}{h} =\frac{2\Omega B_{\rm s}}{c}h
\end{eqnarray}
where $\Omega$ is the spin velocity, and $B_{\rm s}$ is the magnetic field strength on the polar cap.
Therefore, we equivalently treat the inner gap as a parallel-plate capacitor with the voltage being $U'=U$ and the area of the plate, $S$, being the area of the polar cap that
$S\approx 4\pi r_{\rm p}^{2}$ and
\begin{eqnarray}
r_{\rm p}=r_{\ast}\left ( \frac{\Omega r_{\ast}}{c} \right )^{1/2},
\end{eqnarray}
where $r_{\ast}$ is the radius of the pulsar.
Then, the equivalent interval is $h/2$,
the equivalent charge is
\begin{eqnarray}
Q=\frac{\Omega B_{s}hS}{2\pi ck},
\end{eqnarray}
and the equivalent capacitance is
\begin{eqnarray}\label{a0}
C=\frac{Q}{U}=\frac{S\varepsilon_{0}}{2\pi k h},
\end{eqnarray}
where $\varepsilon_{0}$ is the permittivity of vacuum.
Correspondingly, the increase rate of the gap height is exactly the bulk velocity of the charge particles which are flowing outwards.

The charge and discharge in the inner gap
can be equivalent to the charge and break down of the capacitor.
However, such an equality is nonlinear in the sense that both of the capacitance and charge increase with the gap height
until the capacitor is broken down. Moreover, as with the charge of a realistic $RC$ circuit,
we assume that the charge in the inner gap is a transient process (so that equation (\ref{a3}), i.e., the `` dynamic version" of equation (\ref{e2}) holds) and have
\begin{eqnarray}\label{a1}
R\frac{ d(C U)}{dt}+U=U_{\rm max},
\end{eqnarray}
where $R$ is the equivalent constant resistance of the circuit (the whole magnetosphere should be a quasi-static state during a short duration),
\begin{eqnarray}\label{a3}
U=U(t)\approx\frac{\Omega B_{\rm s}}{c}h(t)^{2}
\end{eqnarray}
is the potential across the gap at time $t$, and
\begin{eqnarray}\label{a4}
U_{\rm max}\approx\frac{\Omega B_{\rm s}}{c}h_{\rm max}^{2}
\end{eqnarray}
is the supply voltage (e.g., due to the unipolar induction) with $h_{\rm max}$ being the maximum thickness of the inner gap.
It is worth reminding that $U$ may not be able to reach $U_{\rm max}$ since before that the capacitor may have been broken down.
For example, as illustrated in the inner gap model, the maximum possible potential drop along
any magnetic field line within the polar cap is
\begin{eqnarray}
\Delta V_{\rm max}\approx \frac{\Omega B_{\rm s}}{c} \frac{r_{\rm p}^{2}}{2},
\end{eqnarray}
that is, $h_{\rm max}=0.7 r_{\rm p}$.
However, this potential drop is usually larger than the voltage, $U_{\rm sp}$, required for a spark (see equation (23) in {\cite{1975ApJ...196...51R}}).

Now, equations (\ref{a0})-(\ref{a4}) can be solved as
\begin{eqnarray}\label{a7}
h(t)=\frac{1}{\alpha}\frac{c_{1}\sqrt{\alpha\beta}e^{\sqrt{\alpha\beta}t}-c_{2}\sqrt{\alpha\beta}e^{-\sqrt{\alpha\beta}t}}{c_{1}e^{\sqrt{\alpha\beta}t}+c_{2}e^{-\sqrt{\alpha\beta}t}},
\end{eqnarray}
where
\begin{eqnarray}\label{e11}
\alpha=\frac{kc}{2\Omega r_{\ast}^{3}R\varepsilon_{0}}, \;\;\;\; \beta=\frac{kc}{2\Omega r_{\ast}^{3}R\varepsilon_{0}} h_{\rm max}^{2},
\end{eqnarray}
and $c_{1}$, $c_{2}$ are integration constants.
We set the initial condition as $h(t=0)=0$, and have $c_{1}=c_{2}$ through equation (\ref{a7}). Therefore, the gap hight is given by
\begin{eqnarray}\label{e13}
h(t)=h_{\rm max}\tanh (\sqrt{\alpha\beta}t).
\end{eqnarray}
Through equation (\ref{e13}), the bulk velocity of charge outflow is
\begin{eqnarray}\label{e14}
v_{\rm of}=\frac{dh(t)}{dt}=h_{\rm max}\sqrt{\alpha\beta}[1-\tanh^{2} (\sqrt{\alpha\beta}t)].
\end{eqnarray}
According to equations (\ref{a3}) and (\ref{e13}), we get
\begin{eqnarray}\label{e15}
U_{\rm sp}=\frac{\Omega B_{\rm s}}{c}[h_{\rm max}\tanh (\sqrt{\alpha\beta}\tau)]^{2},
\end{eqnarray}
where $\tau$ is the time that the potential across the gap increases to trigger a spark (then the spark frequency is $\sim 1/\tau$).

Since the multipole component can increase curvature radii of magnetic field lines,
the necessary spatial scale for the reaction $\gamma+\mathbf{B}\rightarrow e^{-}+e^{+}+\mathbf{B}$ will be reduced,
as well as the height and voltage of the gap and the value of $\tau$ (i.e., the spark frequency is increased; see equation (\ref{e15})).
Therefore, the increase of the spark frequency should be the indicator of the emergence of the magnetic multipole component in the polar cap region.

Maybe, such an expectation of the increased spark frequency had been observed in Vela pulsar
(see the evolution of the single pulse profile around 77th pulse shown in Fig. 2 of \citealt{2018Natur.556..219P}).
Interestingly, if the expectation is true,
the detection of an unusually broad radio pulse and a missing next pulse of Vela pulsar \citep{2018Natur.556..219P} can be well understood.
As demonstrated in above, the multipole component will make the gap easier to spark, so if the multipole field extends from the polar cap,
the spark region will be broadened, as will as the pulse profile.
The multipole field lines may bridge the positive and negative charge regions, and the short-lived accelerating field due to the destruction of the force-free condition \citep{1969ApJ...157..869G}
allows electrons to move along the field lines and fill up the gap. So, during this duration, the potential across the gap cannot increase and the spark cannot be triggered.
Only when the net electric field is screened again by the rearrangement of charge particles, can the potential across the gap grow to trigger a spark.
After the nulling pulse, the next two pulses with unexpectedly low linear polarization may also be related to the emergence of the multipole component
since the magnetosphere is altered by the nascent multipole field.

\section{Discussion: possible applications}\label{sec4}
If the prediction of the spark frequency is robustly verified by some observations in the future,
vortex creep model \citep{1975Natur.256...25A}, gap-invoked models of pulsar radio emission \citep{1975ApJ...196...51R,1979ApJ...231..854A} and the model on the origin of magnetic fields of pulsars \citep{1993ApJ...408..194T} are again indicated to be reasonable, and the following applications may be extended to.

\begin{center}
\begin{figure*}
\centering
\includegraphics[scale=.5]{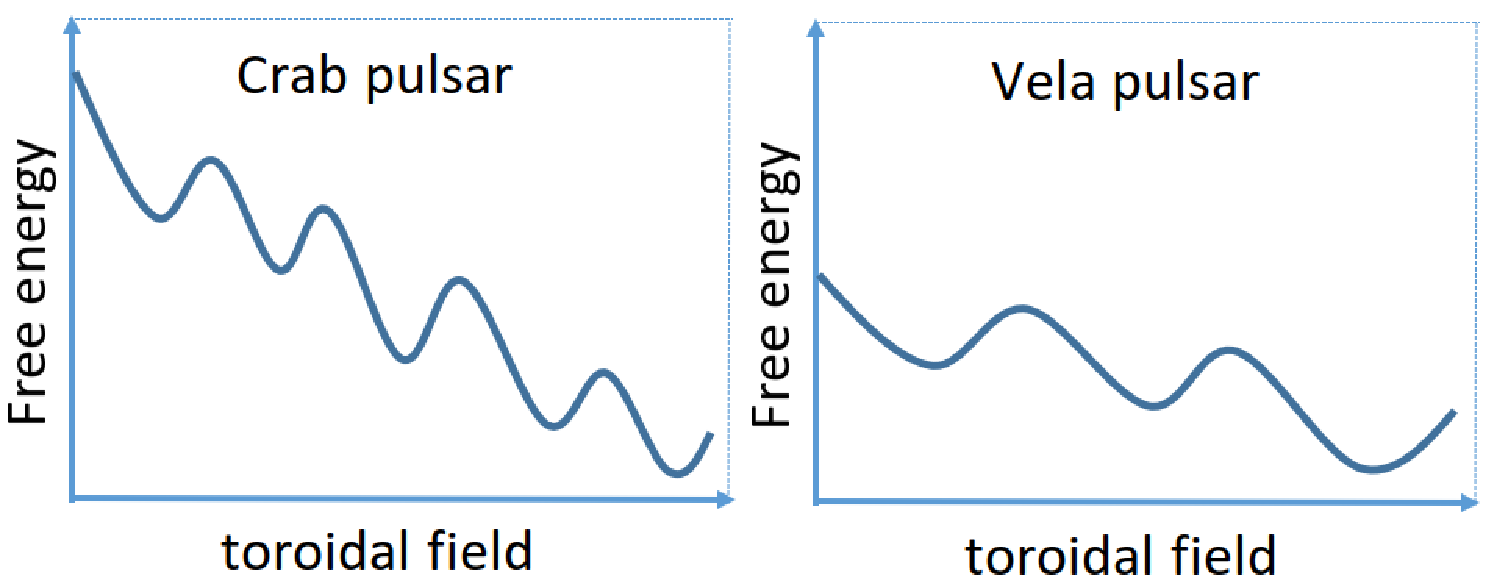}
\caption{
The schematic diagram for illustrating the stabilities of Crab pulsar and Vela pulsar under the perturbation of toroidal magnetic fields.
The strength of the magnetic field decreases along the horizontal axis. Under this scenario, Vela pulsar has a stronger ability to resist perturbations of the toroidal field.
}
\label{f2}
\end{figure*}
\end{center}

\begin{itemize}
  \item[I)] As shown in equation (\ref{e13}), the increase of the potential across the gap follows a hyperbolic tangent function.
  The growth rate will be very slow when the potential is close to the supply voltage, that is, if the value of $U_{\rm sp}$ is close to value of $U_{\rm max}$,
  the corresponding value of $\tau$ will be much larger than that of the case of $U_{\rm sp}\ll U_{\rm max}$.
  Therefore, when the spin period of a radio pulsar, as well as the supply voltage, decays to a certain value,
  the time it takes for the potential across the gap to increase enough to trigger a spark will be longer than the spin period.
  This may be of interest to understand the pulse nulling of some old pulsars \citep{1970Natur.228...42B,1976MNRAS.176..249R,2007MNRAS.377.1383W}
  (another understanding of this phenomenon can refer to \citealt{1986ApJ...300..540D,2004ApJ...600..905L,2005MNRAS.356...59E}).

\item[II)] Observations showed that extinct pulsed radio emission of some magnetars may brighten again after glitches and X-ray bursts  \citep{2006Natur.442..892C,2007ApJ...666L..93C,2010ApJ...721L..33L,2013MNRAS.435L..29S,2020ATel13699....1Z}, and a spin-down glitch \citep{2023NatAs...7..339Y}.
  The cessation of the pulsed radio emission can be explained by the general consideration that the potential across the gap is no longer able to sustain sparks.
  Therefore, the rebrightening of the radio emission should be the consequence of the increased potential (see \citealt{2010MNRAS.408..490M,2015ApJ...799..152L} for another explanation).
  To increase the potential, very curved field lines are required \citep{1993ApJ...402..264C,2000ApJ...531L.135Z}.
  As discussed above, the newborn multipole components due to glitches provide such a condition.
  In addition, according to the above equivalent $RC$ circuit model, curved multiple field lines reduce the potential required to trigger a spark.

\item[III)]
  Crab pulsar and Vela pulsar are both normal radio pulsars and observed glitch events.
  However, post-glitch delayed spin-up\footnote{This name comes from \cite{SLS}.} and persistent shift are only observed in the former (\citealt{1993MNRAS.265.1003L};  the delayed spin-up is also observed in a magnetar \citealt{2022arXiv221103246G}).
  The spin-up indicates the emergence of a net torque (see \citealt{2022arXiv221108151W} for other comments).
  Not that, after the glitch, the previous mechanical equilibrium in the pulsar is destroyed since the current field,
  as well as the distribution of the Lorentz force, is changed.
  The breaking of mechanical equilibrium may prompt the release of the free energy stored in the magentic field.
  For example, a torque, $\propto B_{\rm r}B_{\rm \varphi}$, may act on the pulsar \citep{1999A&A...349..189S} after the glitch,
  such that the toroidal magnetic field untwists and the free energy stored in the toroidal magnetic field can be released.
  Empirically, the toroidal magnetic field of younger Crab pulsar should be stronger than that of older Vela pulsar since the evolution of a
  system always tends to decrease the free energy.
  Therefore, the observed delayed spin-up of Crab pulsar
  could be resulted by the untwisting of the toroidal magnetic field since
  the torque, $\propto B_{\rm r}B_{\rm \varphi}$, can accelerate the spin velocity.
  Meanwhile, the release of the free energy stored in the toroidal magnetic field may enhance the dipole magnetic field and results in the persistent shift.
  This deduction demands that the perturbation of the toroidal magnetic field in the younger Crab pulsar should be more likely to excite a ``transition" (see Figure \ref{f2}).
  The speculation is consistent with experience.
  Once a system reaches a stable state after a long-time evolution,
  the capacity to resist perturbations should be enhanced and closely related to the residual free energy (e.g., the residual toroidal field).
  That is, the younger the pulsar, the easier it is to change.
  For Vela pulsar, there should be only magnetic variation due to the variation of the velocity field and no magnetic variation due to the free energy release when a glitch occurs.
  So, as shown in Figure \ref{f2}, without the release of the free energy,
  the perturbation of the magnetic field cannot be developed into a self-sustaining process,
  and then the altered magnetic field induced by the variation of the velocity field will return to the original state
  when the superfluid vortices repinning to the crust totally (i.e., $\delta\mathbf{v}=0$, $\delta\mathbf{B}=0$, and $\delta\mathbf{j}=0$ in the final state).

\item[IV)] Extra-galactic fast radio burst-like evens from magnetar SGR 1935+2154 were detected to be associated with a glitch \citep{2020Natur.587...59B,2020Natur.587...54C,2022arXiv221103246G} and a spin-down glitch \citep{2023NatAs...7..339Y}.
   These events are poorly understood \citep{2021SCPMA..6449501X}. To explain these events, four questions need to be answered: where the energy comes from, where the charged particles come from, how these particles are accelerated, and how coherent radio emission is produced. As discussed above, glitches will induce the release of the free energy (e.g., magnetic energy and even elastic energy), so the energy reservoir is sufficiently large. The release of the free energy may result in the generation of the new multipole component. If the newborn multipole field lines bridge the positive and negative charge regions on the pulsar surface, the primordial force-free condition \citep{1969ApJ...157..869G} will be briefly violated (during the typical duration $\sim l/c$, where $l$ is the spatial scale of the multipole component). So
  \begin{eqnarray}
  \mathbf{E}+\frac{(\mathbf{\Omega} \times \mathbf{r}\times \mathbf{B})}{c}=0,
  \end{eqnarray}
   changes to
  \begin{eqnarray}
  \mathbf{E}+\frac{[\mathbf{\Omega} \times \mathbf{r}\times (\mathbf{B}+\mathbf{B}')]}{c}\neq 0,
  \end{eqnarray}
  where $\mathbf{\Omega}$ is the spin velocity, and $\mathbf{B}'$ is the strength of the newborn multipole field.
  Once the newborn multipole field, as well as the net electric field $\mathbf{E}'=\frac{1}{c}\left ( \mathbf{\Omega}\times\mathbf{r}\times \mathbf{B}'  \right )$, is strong enough (e.g., $|\mathbf{B}'|\sim 10^{-2}-10^{-4} B_{\varphi}$; note that, $B_{\varphi}$ may be up to $10^{16}\rm G$, \citealt{1993ApJ...408..194T,1995MNRAS.275..255T,1996ApJ...473..322T,2022A&A...666A.138L}),
  electrons in the negative charge region may be accelerated to the required energy along the multipole field lines (e.g., similar to that of electrons accelerated in the gap).
  The fourth question is the most perplexing, even for periodic radio pulses \citep{2018PhyU...61..353B}. Just for the sake of completeness, we suggest such a picture that the accelerated currents along the multipole field lines are cut into bunches by pinch instabilities, and then emit the coherent radio emission via curvature radiation (the question of how the radio emission avoids the scattering and absorption by the plasma near the pulsar surface can referred to \citealt{2019ApJ...879....4W}).
\end{itemize}

\section{Summary}\label{sec5}

In this paper, we show that the commonly invoked magnetic dipole fields of pulsars can not keep steady
even in the presence of toroidal magnetic fields under the general consideration of pulsars,
and multipole components must be generated when the velocity fields in pulsars are varied.
We argue that the increase of the spark frequency of periodic radio pulses is the indicator
of the emergence of the magnetic multipole component.
We present possible explanations for pulse nulling of old pulsars, rebrightning of radio-quiet magnetars, differences between Crab pulsar and Vela pulsar after glitches,
and origin of extra-galactic fast radio burst-like evens from SGR 1935+2154 in terms of the effect of the variation of the velocity fields on pulsar magnetic fields.
While quantitative estimation of the strength of the newborn multipole component is absence (many simulations are required), the logic of these speculations is fluent.

\section{Acknowledgments}
We thank Dr. Weihua Wang for sharing information about glitches with me.
We thank Dr. Zorawar Wadiasingh for reminding me the absorption of the radio emission by magnetospheric plasma.
We thank Dr, Jinchen Jiang for useful discussion.
We thank Pro. Renxin Xu for his help.

\section{Data Availability}
No new data generated.
%(https://pan.baidu.com/s/1q57dU3lX0WYO8DjbRBRwOQ?pwd=xjl0)

%\newpage

\bsp %MNRAS
\label{lastpage} %MNRAS

\begin{thebibliography}{}
\bibitem[Aasi et al.(2014)]{2014ApJ...785..119A} Aasi, J., Abadie, J., Abbott, B.~P., et al.\ 2014, \apj, 785, 119. doi:10.1088/0004-637X/785/2/119
\bibitem[Abbott et al.(2017)]{2017PhRvD..96f2002A} Abbott, B.~P., Abbott, R., Abbott, T.~D., et al.\ 2017, \prd, 96, 062002. doi:10.1103/PhysRevD.96.062002
\bibitem[Anderson \& Itoh(1975)]{1975Natur.256...25A} Anderson, P.~W. \& Itoh, N.\ 1975, \nat, 256, 25. doi:10.1038/256025a0
\bibitem[Arons \& Scharlemann(1979)]{1979ApJ...231..854A} Arons, J. \& Scharlemann, E.~T.\ 1979, \apj, 231, 854. doi:10.1086/157250
\bibitem[Backer(1970)]{1970Natur.228...42B} Backer, D.~C.\ 1970, \nat, 228, 42. doi:10.1038/228042a0
\bibitem[Baym et al.(1969)]{1969Natur.224..673B} Baym, G., Pethick, C., \& Pines, D.\ 1969, \nat, 224, 673. doi:10.1038/224673a0
\bibitem[Beloborodov(2009)]{2009ApJ...703.1044B} Beloborodov, A.~M.\ 2009, \apj, 703, 1044. doi:10.1088/0004-637X/703/1/1044
\bibitem[Beloborodov(2008)]{2008ApJ...683L..41B} Beloborodov, A.~M.\ 2008, \apjl, 683, L41. doi:10.1086/590079
\bibitem[Beskin(2018)]{2018PhyU...61..353B} Beskin, V.~S.\ 2018, Physics Uspekhi, 61, 353. doi:10.3367/UFNe.2017.10.038216
\bibitem[Bochenek et al.(2020)]{2020Natur.587...59B} Bochenek, C.~D., Ravi, V., Belov, K.~V., et al.\ 2020, \nat, 587, 59. doi:10.1038/s41586-020-2872-x
\bibitem[Camilo et al.(2007)]{2007ApJ...666L..93C} Camilo, F., Ransom, S.~M., Halpern, J.~P., et al.\ 2007, \apjl, 666, L93. doi:10.1086/521826
\bibitem[Camilo et al.(2006)]{2006Natur.442..892C} Camilo, F., Ransom, S.~M., Halpern, J.~P., et al.\ 2006, \nat, 442, 892. doi:10.1038/nature04986
\bibitem[Chen \& Ruderman(1993)]{1993ApJ...402..264C} Chen, K. \& Ruderman, M.\ 1993, \apj, 402, 264. doi:10.1086/172129
\bibitem[CHIME/FRB Collaboration et al.(2020)]{2020Natur.587...54C} CHIME/FRB Collaboration, Andersen, B.~C., Bandura, K.~M., et al.\ 2020, \nat, 587, 54. doi:10.1038/s41586-020-2863-y
\bibitem[Deich et al.(1986)]{1986ApJ...300..540D} Deich, W.~T.~S., Cordes, J.~M., Hankins, T.~H., et al.\ 1986, \apj, 300, 540. doi:10.1086/163831
\bibitem[Esamdin et al.(2005)]{2005MNRAS.356...59E} Esamdin, A., Lyne, A.~G., Graham-Smith, F., et al.\ 2005, \mnras, 356, 59. doi:10.1111/j.1365-2966.2004.08444.x
\bibitem[Evans et al.(1980)]{1980ApJ...237L...7E} Evans, W.~D., Klebesadel, R.~W., Laros, J.~G., et al.\ 1980, \apjl, 237, L7. doi:10.1086/183222
\bibitem[Ge et al.(2022)]{2022arXiv221103246G} Ge, M., Yang, Y.-P., Lu, F., et al.\ 2022, arXiv:2211.03246. doi:10.48550/arXiv.2211.03246
\bibitem[Goldreich \& Julian(1969)]{1969ApJ...157..869G} Goldreich, P. \& Julian, W.~H.\ 1969, \apj, 157, 869. doi:10.1086/150119
\bibitem[Haskell \& Melatos(2015)]{2015IJMPD..2430008H} Haskell, B. \& Melatos, A.\ 2015, International Journal of Modern Physics D, 24, 1530008. doi:10.1142/S0218271815300086
\bibitem[Holloway(1973)]{1973NPhS..246....6H} Holloway, N.~J.\ 1973, Nature Physical Science, 246, 6. doi:10.1038/physci246006a0
\bibitem[Hurley et al.(2005)]{2005Natur.434.1098H} Hurley, K., Boggs, S.~E., Smith, D.~M., et al.\ 2005, \nat, 434, 1098. doi:10.1038/nature03519
\bibitem[Jackson(1975)]{1975clel.book.....J} Jackson, J.~D.\ 1975, 92/12/31, New York: Wiley, 1975, 2nd ed.
\bibitem[Kaspi \& Beloborodov(2017)]{2017ARA&A..55..261K} Kaspi, V.~M. \& Beloborodov, A.~M.\ 2017, \araa, 55, 261. doi:10.1146/annurev-astro-081915-023329
\bibitem[Kouveliotou et al.(1998)]{1998Natur.393..235K} Kouveliotou, C., Dieters, S., Strohmayer, T., et al.\ 1998, \nat, 393, 235. doi:10.1038/30410
\bibitem[Levin et al.(2010)]{2010ApJ...721L..33L} Levin, L., Bailes, M., Bates, S., et al.\ 2010, \apjl, 721, L33. doi:10.1088/2041-8205/721/1/L33
\bibitem[Lewandowski et al.(2004)]{2004ApJ...600..905L} Lewandowski, W., Wolszczan, A., Feiler, G., et al.\ 2004, \apj, 600, 905. doi:10.1086/379923
\bibitem[Lin et al.(2015)]{2015ApJ...799..152L} Lin, M.-X., Xu, R.-X., \& Zhang, B.\ 2015, \apj, 799, 152. doi:10.1088/0004-637X/799/2/152
\bibitem[Lin et al.(2022)]{2022A&A...666A.138L} Lin, T., Du, S., Wang, W., et al.\ 2022, \aap, 666, A138. doi:10.1051/0004-6361/202244174
\bibitem[Link \& Epstein(1996)]{1996ApJ...457..844L} Link, B. \& Epstein, R.~I.\ 1996, \apj, 457, 844. doi:10.1086/176779
\bibitem[Lyne et al.(1993)]{1993MNRAS.265.1003L} Lyne, A.~G., Pritchard, R.~S., \& Graham Smith, F.\ 1993, \mnras, 265, 1003. doi:10.1093/mnras/265.4.1003
\bibitem[Moffatt(1978)]{1978mfge.book.....M} Moffatt, H.~K.\ 1978, Cambridge Monographs on Mechanics and Applied Mathematics, Cambridge: University Press, 1978
\bibitem[Morozova et al.(2010)]{2010MNRAS.408..490M} Morozova, V.~S., Ahmedov, B.~J., \& Zanotti, O.\ 2010, \mnras, 408, 490. doi:10.1111/j.1365-2966.2010.17131.x
\bibitem[Palfreyman et al.(2018)]{2018Natur.556..219P} Palfreyman, J., Dickey, J.~M., Hotan, A., et al.\ 2018, \nat, 556, 219. doi:10.1038/s41586-018-0001-x
\bibitem[Peng et al.(2022)]{2022NewA...9001655P} Peng, Q.-H., Liu, J.-J., \& Chou, C.-K.\ 2022, \na, 90, 101655. doi:10.1016/j.newast.2021.101655
\bibitem[Philippov et al.(2020)]{2020PhRvL.124x5101P} Philippov, A., Timokhin, A., \& Spitkovsky, A.\ 2020, \prl, 124, 245101. doi:10.1103/PhysRevLett.124.245101
\bibitem[Qiao \& Lin(1998)]{1998A&A...333..172Q} Qiao, G.~J. \& Lin, W.~P.\ 1998, \aap, 333, 172. doi:10.48550/arXiv.astro-ph/9708245
\bibitem[Radhakrishnan \& Manchester(1969)]{1969Natur.222..228R} Radhakrishnan, V. \& Manchester, R.~N.\ 1969, \nat, 222, 228. doi:10.1038/222228a0
\bibitem[Reichley \& Downs(1969)]{1969Natur.222..229R} Reichley, P.~E. \& Downs, G.~S.\ 1969, \nat, 222, 229. doi:10.1038/222229a0
\bibitem[Ritchings(1976)]{1976MNRAS.176..249R} Ritchings, R.~T.\ 1976, \mnras, 176, 249. doi:10.1093/mnras/176.2.249
\bibitem[Ruderman \& Sutherland(1975)]{1975ApJ...196...51R} Ruderman, M.~A. \& Sutherland, P.~G.\ 1975, \apj, 196, 51. doi:10.1086/153393
\bibitem[Shannon \& Johnston(2013)]{2013MNRAS.435L..29S} Shannon, R.~M. \& Johnston, S.\ 2013, \mnras, 435, L29. doi:10.1093/mnrasl/slt088
\bibitem[Shaw et al.(2018)]{SLS}Shaw B., Lyne A. G., Stappers B. W., et al.\ 2018, MNRAS, 478, 3832. doi:10.1093/mnras/sty1294
\bibitem[Spruit(1999)]{1999A&A...349..189S} Spruit, H.~C.\ 1999, \aap, 349, 189. doi:10.48550/arXiv.astro-ph/9907138
\bibitem[Sturrock(1971)]{1971ApJ...164..529S} Sturrock, P.~A.\ 1971, \apj, 164, 529. doi:10.1086/150865
\bibitem[Thompson \& Duncan(1993)]{1993ApJ...408..194T} Thompson, C. \& Duncan, R.~C.\ 1993, \apj, 408, 194. doi:10.1086/172580
\bibitem[Thompson \& Duncan(1995)]{1995MNRAS.275..255T} Thompson, C. \& Duncan, R.~C.\ 1995, \mnras, 275, 255. doi:10.1093/mnras/275.2.255
\bibitem[Thompson \& Duncan(1996)]{1996ApJ...473..322T} Thompson, C. \& Duncan, R.~C.\ 1996, \apj, 473, 322. doi:10.1086/178147
\bibitem[Usov(1987)]{1987ApJ...320..333U} Usov, V.~V.\ 1987, \apj, 320, 333. doi:10.1086/165546
\bibitem[Wadiasingh \& Timokhin(2019)]{2019ApJ...879....4W} Wadiasingh, Z. \& Timokhin, A.\ 2019, \apj, 879, 4. doi:10.3847/1538-4357/ab2240
\bibitem[Wang et al.(2007)]{2007MNRAS.377.1383W} Wang, N., Manchester, R.~N., \& Johnston, S.\ 2007, \mnras, 377, 1383. doi:10.1111/j.1365-2966.2007.11703.x
\bibitem[Wang et al.(2022)]{2022arXiv221108151W} Wang, W.-H., Ge, M.-Y., Huang, X., et al.\ 2022, arXiv:2211.08151. doi:10.48550/arXiv.2211.08151
\bibitem[Xiao et al.(2021)]{2021SCPMA..6449501X} Xiao, D., Wang, F., \& Dai, Z.\ 2021, Science China Physics, Mechanics, and Astronomy, 64, 249501. doi:10.1007/s11433-020-1661-7
\bibitem[Younes et al.(2023)]{2023NatAs...7..339Y} Younes, G., Baring, M.~G., Harding, A.~K., et al.\ 2023, Nature Astronomy, 7, 339. doi:10.1038/s41550-022-01865-y
\bibitem[Zhang et al.(2000)]{2000ApJ...531L.135Z} Zhang, B., Harding, A.~K., \& Muslimov, A.~G.\ 2000, \apjl, 531, L135. doi:10.1086/312542
\bibitem[Zhang et al.(2020)]{2020ATel13699....1Z} Zhang, C.~F., Jiang, J.~C., Men, Y.~P., et al.\ 2020, The Astronomer's Telegram, 13699
\bibitem[Zhou et al.(2022)]{2022Univ....8..641Z} Zhou, S., G{\"u}gercino{\u{g}}lu, E., Yuan, J., et al.\ 2022, Universe, 8, 641

\end{thebibliography}
\end{document}